%% file: main.tex
\documentclass{article}

\usepackage[utf8]{inputenc}
\usepackage{graphicx}

\usepackage{authblk}
\usepackage{tikz}
\usepackage[normalem]{ulem}

\usepackage{array,makecell}

\usepackage{etoolbox}

\usepackage[hyphens]{url}

\usepackage{todonotes}

\usepackage{fancyhdr}

\usepackage{xcolor}
\usepackage{tcolorbox}
\usepackage{titlesec}
\usepackage{adjustbox}
\usepackage{rotating}
\usepackage{multirow}

\newcommand*\yes{\tikz\draw[black, fill=black](0,0) circle (0.8ex);}
\newcommand*\no{\tikz\draw[black, fill=white](0,0) circle (.8ex);}
\newcommand*\half{
    \begin{tikzpicture}
        \draw[fill=black] (0,0) circle (0.8ex);
        \draw[fill=white] (0,0)-- (90:0.8ex) arc (90:270:0.8ex) -- cycle ;
\end{tikzpicture}}

\definecolor{chaptextbg}{RGB}{220, 220, 220}
\definecolor{chaptext}{RGB}{150, 150, 150}

\newcommand{\printbibliography}[1]{%
  \bibliographystyle{alpha}
  \bibliography{#1}
}

\title{Predicting the Performance of Scientific Workflow Tasks for Cluster Resource Management: \\ An Overview of the State of the Art}

\date{}

\begin{document}


%
%
%
%
%
%
%
%
%
%
%
%
%
%
%
%
%
%
%
%
%
%
%
%
%
%


\include{myChapter}






\end{document}

%% file: myChapter.tex
\newcommand{\hmey}[1]{{\color{blue}[HM: #1]}}

\author[1]{Jonathan Bader}
\author[2]{Kathleen West}
\author[1]{Soeren Becker}
\author[3,4]{Svetlana Kulagina}
\author[3]{Fabian Lehmann}
\author[2]{Lauritz Thamsen}
\author[3,4]{Henning Meyerhenke}
\author[1]{Odej Kao}

\affil[1]{TU Berlin, DOS Group, Berlin, Germany}
\affil[2]{University of Glasgow, School of Computing Science, Glasgow, United Kingdom}
\affil[3]{Humboldt-Universität zu Berlin, Department of Computer Science, Berlin, Germany}
\affil[4]{Karlsruhe Institute of Technology (KIT), Scientific Computing Center, Karlsruhe, Germany}

\maketitle

\begin{abstract}
Scientific workflow management systems support large-scale data analysis on cluster infrastructures.
For this, they interact with resource managers which schedule workflow tasks onto cluster nodes.
In addition to workflow task descriptions, resource managers rely on task performance estimates such as main memory consumption and runtime to efficiently manage cluster resources.
Such performance estimates should be automated, as user-based task performance estimates are error-prone. 

In this book chapter, we describe key characteristics of methods for workflow task runtime and memory prediction, provide an overview and a detailed comparison of state-of-the-art methods from the literature, and discuss how workflow task performance prediction is useful for scheduling, energy-efficient and carbon-aware computing, and cost prediction.

\end{abstract}

\section{Introduction}

When running a scientific workflow on a computing infrastructure, scientists usually have to define task resource consumption limits, such as the available time~\cite{bailey2005user} or memory~\cite{witt2019learning} for task execution.
Setting these resource limits correctly is necessary to enable a successful execution of workflows as exceeding runtime or memory limits results in failures, commonly through errors issued by cluster resource managers.
Resource consumption estimates provided by the scientists are known to be error-prone and inaccurate~\cite{phung2021not,hirales2012multiple,witt2019learning}.
To avoid a failure and timely re-execution of the workflow or parts of it, scientists tend to overprovision resources to ensure that their workflows are fully executed.
The problem is further aggravated as different instances of the same workflow task consume varying amounts of resources~\cite{phung2021not,lehmann2024ponder} when executed on different inputs. 
Assume, for example, that a task unzips its input files. 
Such a task's runtime will heavily depend on the size of the input files.
Hence, scientists have to conservatively configure consumption limits for the largest task instances.

While a failed workflow execution due to an underestimated resource consumption directly affects scientists, an overestimation does so indirectly:
It reduces the number of tasks running in parallel as more cluster resources than necessary are allocated to tasks.
The cluster resource manager ensures that the requested memory is reserved for specific tasks, even if it is not fully used, leaving less memory for other tasks to run as well.
This effectively decreases the level of overall parallelism and, hence, cluster throughput.
Meanwhile, overestimating the runtime of tasks can lead to significantly suboptimal schedules as the scheduler assumes wrong runtimes~\cite{ramirez2017adaptive, ilyushkin2018impact}.

Predicting the performance of workflow tasks to set their resource consumption limits automatically can relieve scientists of the need to define limits manually.
State-of-the-art approaches often focus on either runtime or memory prediction, as these are key examples of performance prediction where underestimation leads to task termination in resource managers~\cite{ramirez2017adaptive,witt2019predictive,tovar2022dynamic}.
Many approaches use machine learning to predict the performance of workflow tasks. 
Some of these approaches rely on historical data to model a task's performance~\cite{hilman2018task,witt2019learning}, while other approaches learn during the workflow execution and incrementally update their model online~\cite{witt2019feedback,tovar2017job}.
In addition, some approaches use profiling to predict the performance of workflow tasks before the workflows are executed~\cite{bader2022lotaru, bader2024lotaru}.
The approaches also differ in their ability to predict performance for different types of infrastructure, such as homogeneous or heterogeneous infrastructures, their used prediction model, or their model input.

\begin{figure}[]
\includegraphics[width=1.0\columnwidth]{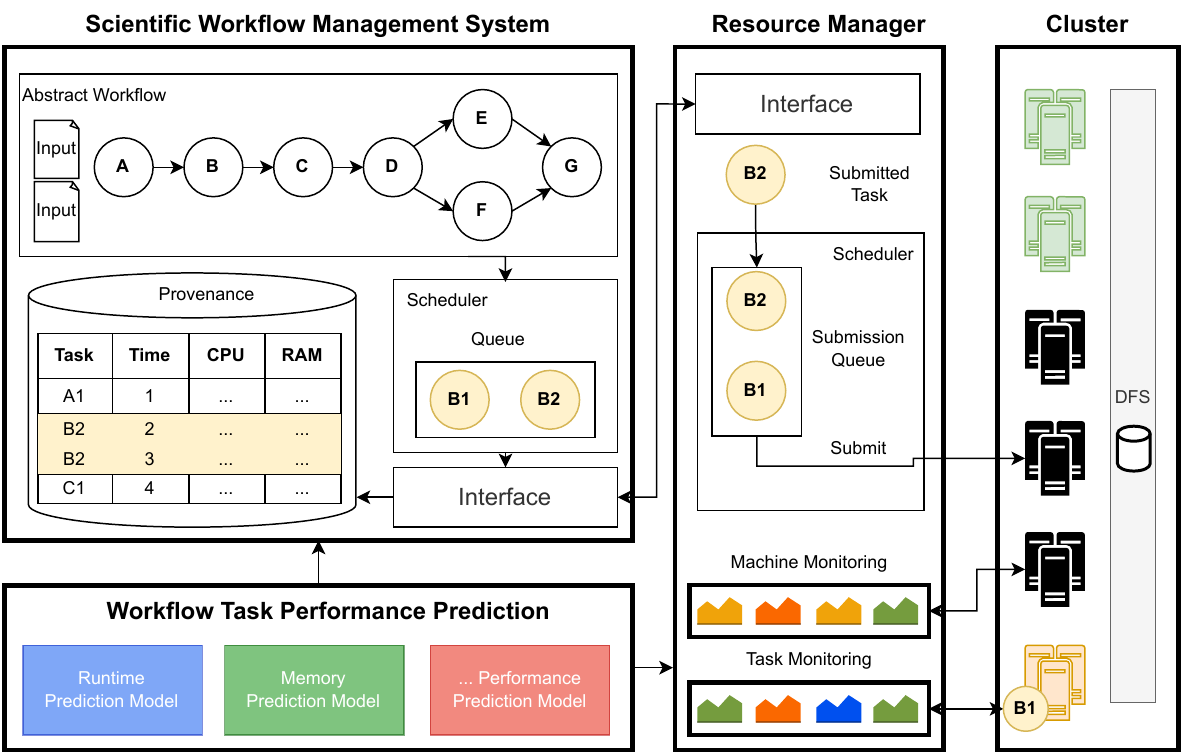}
	\caption{
 Interaction between a scientific workflow management system, a resource manager, and task performance prediction models. }
	\label{fig:book_interaction_sw_executon_environment}
\end{figure}

Figure~\ref{fig:book_interaction_sw_executon_environment} provides a high-level overview of the interactions among components within the scientific workflow execution environment.
The Scientific Workflow Management System (SWMS) parses the abstract workflow and coordinates task submission to the resource manager, which then schedules and allocates tasks onto the cluster infrastructure, concurrently monitoring their resource usage.
Workflow task performance prediction employs predictive models to generate runtime, memory, and other resource-related estimations for individual tasks.
These predictions are accessible to both the SWMS and the resource manager.
Utilizing predictions at the SWMS level offers the advantage of leveraging workflow graph information, which is typically unavailable to the resource manager.
Conversely, the resource manager benefits from a global view, enabling it to optimize resource allocation decisions across multiple workflows and users.
Recent work proposes unified solutions that leverage information from both components, enabling direct integration of task performance predictions~\cite{ahn2020flux,lehmann2023workflow}.

In this chapter, we give an overview of state-of-the-art workflow task performance prediction methods, compare them, and discuss use cases as well as ongoing research.
In particular, we compare methods that focus on \textbf{workflow task runtime} and \textbf{workflow task memory} prediction.
That is, we do not include methods that only predict the total execution time or total memory consumption of entire workflow applications, as task-level predictions serve as fine-grained foundation for workflow-level predictions.
Likewise, we do not include methods that are not specific to scientific workflows.
The main contributions of this chapter are:

\begin{itemize}
    \item We identify key characteristics of workflow task runtime (Section~\ref{sec:runtimepredictioncharacteristics}) and memory prediction methods (Section~\ref{sec:memorypredictioncharacteristics}), covering general aspects such as code availability, model-specific criteria like the prediction method and applicability to heterogeneous infrastructures, as well as model inputs and evaluation settings.
    \item We provide an overview and a detailed comparison of state-of-the-art workflow task runtime (Section~\ref{sec:RuntimePred}) and memory prediction methods (Section~\ref{sec:MemoryPred}), analyzing nineteen state-of-the-art papers based on the previously defined characteristics and identifying shortcomings while highlighting future research directions.
    \item We exemplify how workflow task performance prediction is useful for resource management, namely scheduling, energy-efficient and carbon-aware computing, and cost prediction, and present recent research in each of these fields while also highlighting limitations (Section~\ref{sec:PPforRM}).
\end{itemize}

The remainder of this chapter is structured as follows. 
Section~\ref{sec:RuntimePred} describes state-of-the-art workflow task runtime prediction methods and compares them, while Section~\ref{sec:MemoryPred} describes state-of-the-art workflow task memory prediction methods.
Section~\ref{sec:PPforRM} presents applications of workflow task performance prediction for resource management and discusses the shortcomings of state-of-the-art methods.
Lastly, Section~\ref{sec:conclusion} concludes the chapter.


\begin{sidewaystable}
    \centering
    \caption{Overview of state-of-the-art methods for workflow task runtime prediction.}
\begin{tabular}{|l||c|c||c|c|c|c||c|c|c|c||c|c|c|}
\hline
& \multicolumn{2}{c||}{General} & \multicolumn{4}{c||}{Model} &
\multicolumn{4}{c||}{Model Input} &
\multicolumn{3}{c|}{Evaluation} \\
\hline
\rotatebox{90}{Publication} & \rotatebox{90}{Code Available} & \rotatebox{90}{Domain Specific} & \rotatebox{90}{Prediction Method} & \rotatebox{90}{Training Time} & \rotatebox{90}{Heterogeneous Infrastructure} & \rotatebox{90}{GPU Support} & \rotatebox{90}{Hardware Capacity} & \rotatebox{90}{Hardware Performance} & \rotatebox{90}{Task Input Size} & \rotatebox{90}{Task Resource Usage} & \rotatebox{90}{Workflow System} & \rotatebox{90}{\# Workflows} & \rotatebox{90}{Experiment Type} \\
\hline
\cite{matsunaga2010use} & \no & \no & Diverse & Offline &  \yes & \no &  \yes &  \yes &  \yes & \no & - & - & Sim. Real \\ \hline
\cite{malik2013execution} & \no & \no & NN & Offline &  \yes & \no &  \yes &  \yes &  \yes &  \yes & Askalon & 3 & Sim. Real \\ \hline
\cite{da2013toward} & \no & \no & Regr. \& Clust. & Online & \no & \no & \no & \no &  \yes & \no & Pegasus & 3 & Sim. Real \\ \hline
\cite{da2015online} & \no & \no & Regr. \& Clust. & Online & \no & \no & \no & \no &  \yes & \no & Pegasus & 5 & Sim. Real \\ \hline
\cite{pham2017predicting} & \no & \no & Regr. & Offline &  \yes & \no &  \yes &  \yes &  \yes &  \yes & - & 4 & Sim. Syn. \\ \hline
\cite{nadeem2017modeling} & \no & \no & NN & Offline & \no & \no & \no & \no &  \yes & \no & - & 3 & Sim. Real \\ \hline
\cite{hilman2018task} & \no & \no & NN & Online &  \yes & \no &  \yes &  \yes &  \yes &  \yes & Diverse & 2 & Sim. Real \\ \hline
\cite{bader2022lotaru} &  \yes & \no & Regr. & Pre &  \yes & \no &  \yes &  \yes &  \yes &  \yes & Nextflow & 5 & Sim. Real \\ \hline
\cite{huang2023cloudprophet} & \no & \no & NN & Offline &  \yes & \no &  \yes &  \yes & \no &  \yes & - & - & Sim. Real \\ \hline
\cite{bader2024lotaru} &  \yes & \half & Regr. & Pre &  \yes & \no &  \yes &  \yes &  \yes & \no & Nextflow & 5 & Sim. Real \\ \hline

\multicolumn{4}{l}{\yes Yes | \no  No | \half Partially}\\ 
\end{tabular}
\label{tab:runtime}
\end{sidewaystable}

\section{Workflow Task Runtime Prediction}
\label{sec:RuntimePred}

In this chapter, we explore the features of runtime prediction methods for workflow tasks and offer a detailed overview and thorough comparison of the latest task runtime prediction techniques found in the literature.
Table~\ref{tab:runtime} gives an overview.

\subsection{Characteristics}
\label{sec:runtimepredictioncharacteristics}

In order to provide a high-level overview and compare methods for predicting the runtime of workflow tasks, we introduce a set of characteristics divided into four main areas: general, model, model input, and evaluation.

\paragraph{General:} First, we describe the general characteristics of workflow task runtime prediction methods, including information about the \emph{publication year}, \emph{public code availability}, and whether the method is \emph{domain-specific} or can be applied to all types of workflows. 

\paragraph{Model:} The model characteristics describe essential workflow task runtime prediction models' attributes.
The \emph{prediction method} attribute describes the models' underlying prediction method, such as a regression or a neural network.
The \textit{training time} attribute describes when the model is trained, either offline on historical data, online during workflow execution, or pre-workflow execution using profiling without historical data.
The \emph{heterogeneity} attribute shows whether a method is applicable to heterogeneous infrastructures or restricted to homogeneous ones.
With the recent popularity of computing on GPUs, we have added a \emph{GPU support} attribute as additional information to show whether a prediction method lends itself to running workflow tasks on both CPUs and GPUs.

\paragraph{Model Input:} The model input shows the data used to perform the runtime prediction.
The first attribute is the hardware capacity that shows whether static hardware information, such as the number of CPU cores, amount of memory, or disk sizes, is incorporated. 
Complementary, the \emph{hardware performance} characteristic considers performance metrics in the model input, such as processor speed or I/O capabilities.
An important feature for workflow task runtime prediction is the \textit{task input data size}, which can be the general input size or a placeholder, for instance, bytes read.
Finally, we include \emph{task resource utilization} as an attribute and analyze whether the models use task resource utilization, such as CPU or memory utilization, as an input.

\paragraph{Evaluation:} The attributes under the evaluation category show how the proposed methods were evaluated.
The first attribute, \emph{workflow system}, shows which workflow system was used to either run the experiments or gather the traces, followed by the \emph{number of workflows} attribute, aiming to quantify the scope.
The \emph{experiment type} shows if the method was evaluated on real infrastructures, simulated with traces gathered from real executions, or simulated with synthetic traces.

\subsection{Offline Workflow Task Runtime Prediction}

Offline predictors build their model based on historical data and before the workflow execution is started, without the capabilities to improve the model online during workflow execution.  

Matsunaga~et~al.~\cite{matsunaga2010use} perform offline runtime prediction and evaluate several machine learning approaches for their ability to predict task runtime.
Their method can be applied to heterogeneous infrastructures as it incorporates static and performance characteristics of the hardware.
They conducted their evaluation on two bioinformatic tools, including an analysis of the impact of individual features on prediction accuracy, arguing for including as many features as possible and letting the algorithm decide on the specific selection. 
The authors also show that different algorithms perform better for different setups, i.e., the effectiveness of an algorithm depends on the task and its training data.

Malik~et~al.~\cite{malik2013execution} leverage runtime provenance data to train a multilayer perceptron (MLP)-based neural network that predicts task execution times on grid infrastructures. 
Model input features include static and dynamic environment features such as a machine's CPU speed or free memory, pre-execution features such as input size, and execution features such as task resource usage.
Their model is trained offline, using a feature set optimizer, a model space analyzer, an MLP generator, and a model trainer component.
The authors evaluate their method with three workflows, MeteoAG, Invmod, and Wien2K, using Askalon as the workflow system, and show a prediction error between 17\% and 26\%.

Pham~et~al.~\cite{pham2017predicting} present an offline task runtime prediction method for cloud environments using a two-stage approach that first predicts task resource usage, followed by a second stage that uses the output of the first stage, the workflow input data, and static and dynamic hardware information for a regression to predict execution time on a target machine.
Their prediction model distinguishes between pre-runtime parameters, e.g., workflow input data or VM types, and runtime parameters, such as CPU, memory, I/O operations, and bandwidth. 
In the first stage, pre-runtime parameters are considered to derive the runtime parameters for the execution.
In the second stage, the task execution time on a target VM is predicted with a regression model using the output data from the first stage, the workflow input data, and the VM specifications.
Their method is applicable to heterogeneous infrastructures and evaluated with generated data from four different workflows and diverse scientific workflow management systems. 
 
Nadeem et al.~\cite{nadeem2017modeling} present a method for offline prediction of task execution times in grid environments using a radial basis function neural network.
The learning model incorporates four specific types of information: a) workflow structure, including aspects like workflow name or dependency flow; b) application details, such as executable names or file sizes; c) execution environments, covering factors like grid locations or submission time; and d) resource state, including metrics like the number of jobs in the queue or jobs running. 
However, it does not account for hardware characteristics; hence it is not able to predict runtimes for heterogeneous infrastructures.
For their evaluation, they used generated data from four different workflows and various scientific workflow management systems. 
Similar to Matsunaga~et~al.~\cite{matsunaga2010use}, their evaluation outcome shows which features have the greatest impact on the predicted runtime and which ones can be left out.

Huang~et~al.~\cite{huang2023cloudprophet} propose CloudProphet, a method for predicting task execution times for public cloud infrastructures.
CloudProphet operates from the view of a data center operator, also assuming that the VM where the task is running, is a black box.
Thus, in a first step, the application type is identified based on task resource consumption, using a Dynamic Time Warping algorithm to normalize the temporal difference in the training traces.
Next, highly correlated metrics are selected.
Then, a neural network predicts the performance of the task using the highly correlated metrics as input.
In addition, the workload level is predicted using a second neural network that outputs the task's workload level.
A third neural network is then used that takes the output of the first as input and predicts the performance baseline that can be used to calculate the performance degradation.
Their evaluation uses five cloud benchmarks from CloudSuite and shows similar prediction accuracy to Pham~et.~al.~\cite{pham2017predicting} and Bader~et~al.~\cite{bader2022lotaru}.

\subsection{Online Workflow Task Runtime Prediction}
\label{subsec:onlineRuntime}

Da~Silva~et~al.~\cite{da2013toward} predict task resource consumption, such as runtime, disk space, and memory consumption, for tasks in scientific workflows.
Based on monitoring tools and historical data, they apply a regression tree for resource prediction.
Beforehand, they identify data subsets with a high correlation by using density-based clustering.
Then, they predict the expected resource usage for correlated data points based on the ratio in this specific cluster.
In the case of uncorrelated data, they predict the mean runtime of previous executions.
The authors evaluate their method with three workflow traces gathered on real infrastructures using the Pegasus workflow management system.

In an extension of their work, da Silva et al.~\cite{da2015online} test a normal and a gamma distribution for uncorrelated data and sample in the case of statistical significance.
They also extend their evaluation with two new workflow traces gathered from real executions.

Hilman~et~al.~\cite{hilman2018task} employ an online incremental learning method utilizing long short-term memory networks (LSTMs) for predicting task runtimes in heterogeneous cloud environments.
Their method uses two categories of metrics: pre-runtime and during runtime. 
The pre-runtime metrics encompass details of the task, virtual machine (VM) specifications, and the time of task submission. 
In contrast, the runtime metrics include parameters like CPU usage, memory consumption, and I/O operations, represented as historical time-series data. 
This data is continuously augmented online, during the workflow execution, and is utilized to train and refine the model after task completion.
The method is evaluated on two workflows with generated tasks.

\subsection{Pre-Execution Workflow Task Runtime Prediction}

In our own work, we proposed the Lotaru method~\cite{bader2022lotaru, bader2024lotaru}, which predicts workflow task runtimes for heterogeneous infrastructures without using historical data and before the workflow is executed on the target infrastructure.
Lotaru quickly profiles the local machine and the target infrastructure using a set of microbenchmarks, then runs the workflow with downsampled data on the local machine to collect task metrics.
Using the hardware performance, task input, and resource usage data, tasks are assessed as CPU- or I/O-intensive, and a Bayesian linear regression model is trained to predict the runtime on target machines.
For evaluation, five real-world workflows from the nf-core repository~\cite{ewels2020nf} were executed on six different node types using the Nextflow workflow management system, which generated experimental traces.

\subsection{Discussion}

Table~\ref{tab:runtime} provides an overview of the state-of-the-art workflow task runtime prediction methods presented in the literature. 

Of the eleven methods, only two methods provide their source code, experimental setup scripts, or configurations as open source.
With limited access to the source code, the reproducibility of the results and also their applicability for real-world scientific workflows is limited.

One of the methods is partially domain-specific for bioinformatics and uses data attributes to downsample the input for profiling.
The other methods are generally applicable, but are often evaluated only for a single domain.

For the machine learning prediction method, we can see that the majority uses neural networks and regression-based models.

Five methods predict task runtimes offline, three online, and two before workflow execution using profiling.
Thus, there are different methods for each part of the workflow execution cycle.

Seven methods are able to predict runtimes for heterogeneous environments such as clouds, but none of these methods consider the use of GPUs.

The model inputs differ significantly between the different methods. 
All but one method make use of the task input size. 
Three methods furthermore only make use of the input size, while all other methods consider multiple factors for predictions, including hardware performance and capacity as well as task resource usage.

All methods are evaluated in a simulation setup, mostly using traces from real workflow executions on real infrastructures, with the most commonly used workflow systems being Pegasus and Nextflow.
Evaluating these methods by integrating and evaluating them in real execution environments is an open research point.
This would, for example, allow to study the two-way interactions between online task runtime predictors and schedulers that make use of predictions that can occur in real systems.


\begin{sidewaystable}
    \centering
    \caption{Overview of state-of-the-art workflow task memory prediction methods.}
\begin{tabular}{|l||c|c||c|c|c|c|c||c|c|c|}
\hline
& \multicolumn{2}{c||}{General} & \multicolumn{5}{c||}{Model} &
\multicolumn{3}{c|}{Evaluation}
 \\ \hline
\rotatebox{270}{Publication} & \rotatebox{270}{Code Available} & \rotatebox{270}{Domain Specific} & \rotatebox{270}{Prediction Method} & \rotatebox{270}{Online/Offline} & \rotatebox{270}{Heterogeneous Infrastructure} & \rotatebox{270}{Error Handling} & \rotatebox{270}{Task Input Size} & \rotatebox{270}{Workflow System} & \rotatebox{270}{\# Number Workflows} & \rotatebox{270}{Experiment Type} \\
\hline
\cite{da2013toward} & \no & \no & Regr. \&  Clust. & Online & \no & \no & \yes & Pegasus & 3  & Sim.  \\ \hline
\cite{da2015online} & \no & \no & Regr. \&  Clust. & Online & \no & \no & \yes & Pegasus & 5  & Sim.  \\ \hline
\cite{tovar2017job} & \yes & \no & Analytical & Online & \no & \yes & \no & Makeflow & 3 & Sim. \\ \hline
\cite{witt2019learning} & \yes & \no & Regr. & Offline & \no & \yes & \yes & IceProd & 1 & Sim.  \\ \hline
\cite{witt2019feedback} & \no & \no & Regr. & Online & \no & \yes & \half & Pegasus & 5 & Sim Syn. \\ \hline
\cite{bader2022leveraging} & \no & \no & RL & Online & \no & \yes & \no & Nextflow & 5 & Real  \\ \hline
\cite{tovar2022dynamic} & \yes & \yes & Analytical & Online & \no & \yes & \no & Coffea & 1 & Real  \\  \hline
\cite{baderPredictDynamicMemoryRequ2023} & \yes & \no & Time Series Regr. & Online & \no & \yes & \yes & Nextflow & 2 & Sim. Real \\ \hline
\cite{bader2024Sizey} & \yes & \no & Multiple & Online & \no & \yes & \yes & Nextflow & 6 & Sim. Real \\ \hline
\cite{lehmann2024ponder} & \yes & \no & Rule \& Regr. & Online & \no & \yes & \yes & Nextflow & 4 & Real \\ \hline
\cite{bader2024ks+} & \yes & \no & Time Series Regr. & Online & \no & \yes & \yes &  Nextflow & 2 & Sim. Real \\ \hline
\multicolumn{4}{l}{\yes Yes | \no  No | \half Partially}\\ 
\end{tabular}
\label{tab:memPred}
\end{sidewaystable}

\section{Workflow Task Memory Prediction}
\label{sec:MemoryPred}

In this section, we cover the characteristics of workflow task memory prediction methods and provide a detailed overview and comprehensive comparison of state-of-the-art memory prediction methods from the literature.
Table~\ref{tab:memPred} gives a structured overview of existing memory prediction methods.

\subsection{Characteristics}
\label{sec:memorypredictioncharacteristics}

In this section, we describe the key features, which we divide into three areas: General, Model, and Evaluation. 

\paragraph{General:} Again, the general characteristics include information about the \emph{year of publication} and \emph{code availability}.
It also includes a classification of general or \emph{domain-specific} applicability, as we observed that some methods require certain data properties that are only fulfilled for certain domains or use cases.

\paragraph{Model:} The model characteristics include details about the underlying \emph{prediction model}, machine learning or statistical models such as regression trees, linear regression methods, reinforcement learning, or ensemble methods that use multiple machine learning models. 
The model feature also encompasses the \emph{training time}, whether memory is predicted from historical traces (i.e., offline) or online during execution and if \emph{prediction errors} are handled for subsequent predictions.
Furthermore, we examined whether machine \emph{heterogeneity} is considered, such as different processors, memory configurations, and memory access architectures (e.g., UMA vs. NUMA).
We do not divide this characteristic further, as we found no support in existing state-of-the-art methods.
Finally, we analyze whether the prediction model assumes a correlation between \emph{input data size} and memory usage.

\paragraph{Evaluation:} Again, we examine whether the \emph{experiment type} uses real executions, a simulation on real data, or a simulation on synthetic data.  
We also look at the \emph{number of workflows} used and the \emph{workflow management system} used either to execute the workflow or to collect traces.

\subsection{Offline Task Memory Prediction}

Witt~et~al.~\cite{witt2019learning} propose a method that optimizes the resource wastage instead of the prediction error. 
The authors show a correlation between the size of the input data and the memory consumption of a task and accordingly develop a linear prediction model that incorporates the input sizes as a feature.
Their method doubles the predicted memory on a task failure.
The method is evaluated using a simulation with the traces of a Neutrino Observatory experiment log.
The authors' method achieves less wastage than user estimates and the baseline of Tovar~el~al.~\cite{tovar2017job}.
In their experiments, they further investigate different failure handling strategies, i.e., how to handle task failures due to memory underestimation, and show a significant impact on wasted memory resources, but also that there is no best strategy for all kinds of tasks.

\subsection{Online Task Memory Prediction}

The method developed by da~Silva~et~al.~\cite{da2013toward} and its extension~\cite{da2015online} can also be applied to memory prediction for correlated data points based on the ratio in this specific cluster.
We provided a comprehensive overview of this method in the previous section~\ref{subsec:onlineRuntime}.

Tovar~et~al.~\cite{tovar2017job} introduce a strategy for predicting task memory in scientific workflows. 
The authors develop a predictive model that estimates the peak memory demand of a given task using an online analytical approach during workflow execution. 
The model is adjustable, focusing on either maximizing throughput or minimizing resource wastage. 
Their optimization is based on the slow-peaks model, which assumes the scenario where task failures occur towards the end of the execution. 
The authors suggest using a two-step policy that first allocates the predicted amount of memory, followed by allocating the maximum available resources in case a task execution fails. 
While the authors acknowledge the potential for a more complex multi-phase failure handling policy, they note it requires more in-depth investigation. 
Their method is evaluated by a simulation using traces from three workflow executions with the Makeflow workflow management system, showing an overall reduction in memory wastage and an increase in throughput.

Tovar~et~al.~\cite{tovar2022dynamic} present a domain-specific solution to predict memory for task executions and splitting tasks into smaller slices.
The first five executed tasks will be assigned the maximum memory of a node.
Afterward, the actual used memory is monitored, and the maximum ever observed is assigned plus a small margin.
When a task fails, it is split into two equally sized tasks and resubmitted, leading to potentially multiple subsequent retries and splits.
However, the authors also mention that the possibility of task splitting is due to the nature of independent events that comprise the tasks, i.e., the domain knowledge.
Initially, the task size is determined by the linear correlation between the chunk size of a task and its resource usage.
Task splitting has also been proposed in the field of genomics, but is not generally applicable to black-box tasks~\cite{mohammadi2024optimizing}.
Their evaluation uses the coffea framework and runs the TopEFT workflow on a cluster of 40 commodity nodes.
The results show the efficiency of their method, reducing wasted memory resources.

Witt~et~al.~\cite{witt2019feedback} present a feedback-based peak memory allocation method.
The authors propose two different predictors.
The first predictor is a linear regression model that uses the sum of the task's input files' size as a feature to predict peak memory.
They dynamically offset their prediction to avoid an underallocation.
For example, the ``LR~mean~±'' strategy incorporates the standard deviation as an offset, while the ``LR~mean~-'' method accounts solely for negative prediction errors.
As a second predictor, they use the percentile of historical peak memory usage, such as the median or the 99\textsuperscript{th} percentile.
In this case, no offset is selected.
For failed task executions, the previously predicted peak memory will be multiplied by a factor of two.
The methods are evaluated using the DynamicCloudSim simulation environment with five synthetic traces from the Pegasus Workflow Generator.
The authors also examine the relationship between wasted memory and scheduling policies and show that the two mutually influence each other.

We~\cite{bader2022leveraging} present two reinforcement learning methods that predict the peak memory requirements of tasks.
Our first method is based on gradient bandits that must learn a preference for a particular memory configuration.
The agent's reward function includes a penalty for memory underallocations to discourage the agent from choosing such actions.
Their failure strategy ensures that no allocations smaller than the failed one are chosen.
The second method uses Q-learning agents that increase, decrease, or maintain memory allocation. 
The agents' reward function includes an artificial minimum and maximum amount of memory to allocate and discourages insufficient memory allocations.
We evaluate the reinforcement learning approaches using five workflows from the nf-core repository and ran them on real infrastructures, showing that the methods are able to reduce memory wastage compared to the workflow defaults.

Lehmann~et~al.~\cite{lehmann2024ponder} propose a memory sizing strategy called Ponder to dynamically decide whether a linear relationship between a task's data input size and the memory consumption exists.
Ponder aims to reduce the chance of out-of-memory failures and uses a rule-based method to predict a task's memory.
If a relationship is detected, linear regression is applied; otherwise, the highest observed memory value is used. 
The method also adjusts predictions to ensure they are reasonable, particularly when they exceed historical values.
Additionally, Ponder includes a weighted offset and a static offset to account for variability in memory consumption with the same input data and task across different runs.
The authors evaluate their approach for four different nf-core workflows on Kubernetes.

We~\cite{bader2024Sizey} recently introduced an online method that uses multiple machine learning methods and dynamically selects the best-performing one.
During workflow execution, the model is continuously retrained and updated to incorporate metrics from recent task executions.
A novel resource allocation quality score is used to continuously assess the goodness of a model's prediction, which is based on accuracy and efficiency.
We use not one but many offsetting techniques and dynamically select the one that would have caused the lowest resource wastage.
The method is evaluated on traces of six real-world workflows gathered with the workflow management system Nextflow.
The experimental results significantly outperform existing state-of-the-art peak memory prediction methods from the literature.

We~\cite{baderPredictDynamicMemoryRequ2023,bader2024ks+} also propose a memory prediction approach that uses time series monitoring data in order to provide a dynamic memory prediction, i.e., a memory prediction over time.
Due to the dynamic approach, the method first predicts the runtime of a workflow task by training a linear model and offsets the predicted time.
Then, the expected runtime of the task is segmented according to predefined values.
Now, for each of the segments, a linear regression model is trained to predict the maximum memory usage for each segment.
The prediction is offset for each segment to avoid memory underallocation.
Since our model has two aspects, runtime and memory, we discuss and propose different error-handling methods.
For example, the selective retry strategy, which adjusts only the failed segment, or the partial retry strategy, which adjusts all segments from the one that failed.
The method is evaluated in a simulation using traces from two real-world workflows from the publicly available nf-core repository and outperforms state-of-the-art methods in terms of reducing memory wastage.

\subsection{Discussion}

Table~\ref{tab:memPred} gives an overview of eleven state-of-the-art methods for predicting memory usage for workflow tasks from the literature.
We divided the characteristics we included in our comparison into three categories: General, Model, and Evaluation.

Of the eleven methods presented, seven make their source code publicly available.
Compared to workflow task runtime prediction methods from the literature, this is a higher percentage.
This makes it easier to use existing methods as baselines for newer approaches if prototype implementations are readily available. 

One of the presented methods can only be applied to a specific domain, all others are generally applicable.
The prediction models are very heterogeneous and use different machine learning methods.

One method is applied offline, while ten methods run online.
This is a clear difference to task runtime prediction methods, where only a couple of methods learned prediction models during the runtime of workflows.  

Strikingly, none of the methods considers a heterogeneous infrastructure.
It can be assumed that this is the case because, from our experience, the memory consumption of the same workflow task on different machines is often quite stable, especially when using the same operating system and the same underlying processor architecture, unlike the runtime of the workflow task, so that there is less need to evaluate methods across different machines.

Nine of the eleven methods perform error handling, i.e., they adjust the prediction after a task failure.
Many of the papers emphasized that the use of error handling has a significant impact on wasted memory resources and is thus almost as important as the initial memory prediction.

Eight methods are evaluated using simulations, while three evaluate their method on real systems.
We believe that most methods are simulated for practical reasons: The execution of large workflows is time-consuming, and re-running parts due to memory failures further increases the experiment time while consuming valuable cluster resources and energy.
In addition, simulations make it easy to try out a variety of error-handling strategies and method configurations.
However, once a well-working method has been found, it would be beneficial to evaluate the methods in real-world systems, as this would increase the integration in real scientific workflow management systems.

Overall, Pegasus was a commonly used workflow management system for evaluation in older publications, while newer publications often use Nextflow.

\section{Applications of Task Performance Prediction for Scientific Workflow Resource Management}
\label{sec:PPforRM}

\begin{figure}[]
\includegraphics[width=1.0\columnwidth]{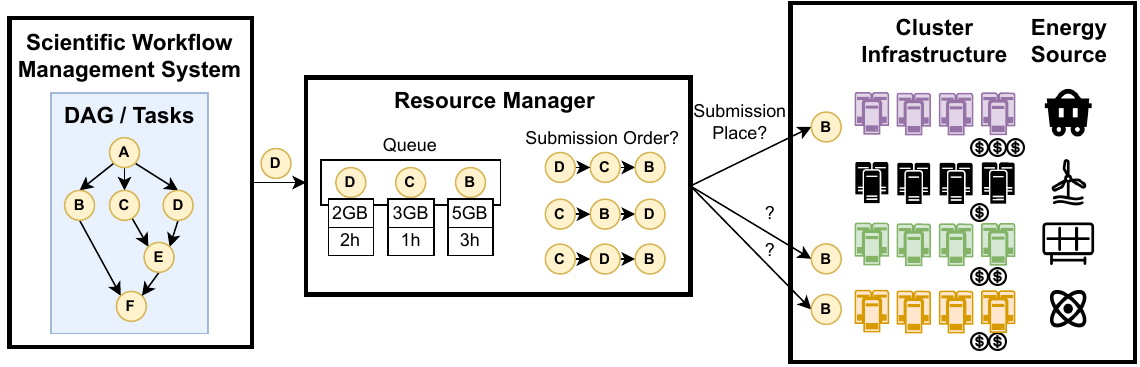}
	\caption{
 Typical execution environment where task performance predictions are used for scheduling, energy-efficient and carbon-aware execution, and cost prediction. The cluster can take different forms, such as an on-premise system, a virtual cluster in the cloud, or a hybrid solution. }
	\label{fig:overview_resource_management}
\end{figure}

Many advanced resource management methods make use of workflow task resource predictions.
Prominent examples include workflow scheduling methods that aim to minimize overall workflow execution time (modeled by the makespan)~\cite{topcuoglu2002performance,samadi2018eheft}, methods for energy- and carbon-efficient workflow execution~\cite{durilloMultiobjectiveEnergyefficientWorkflow2014b,souza2023ecovisor}, and methods that aim to predict and optimize the cost of executing workflows~\cite{alkhanak2016cost,rosa2021computational}.
Figure~\ref{fig:overview_resource_management} provides an overview of a typical execution environment for scientific workflows and shows how resource prediction can be employed.
When scheduling to minimize the makespan, the resource manager uses predicted resources, such as memory and runtime, to determine the task submission order and the best node for each task.
Energy-efficient and carbon-efficient workflow execution incorporates information about the energy efficiency of available cluster machines and the availability of low-carbon energy. 
Finally, cost prediction considers the price of using cluster machines and uses resource estimates to select the most cost-effective machine for a task.
In the following, we describe each of these resource management methods in more detail and summarize relevant work.

\subsection{Workflow Scheduling:}

\paragraph{Introduction:} Scheduling tasks has been a central topic in workflow research for decades~\cite{heft,weske1996scientific}. 
For large-scale workflows, which are often executed on parallel and distributed platforms, scheduling involves deciding where and when to execute each task within the workflow. 
These decisions are guided by specific objectives, typically aiming to optimize performance or resource utilization.

Scheduling approaches can typically be categorized into online and plan-based strategies. 
Online scheduling assigns tasks to resources as they become available, which provides flexibility but often limits global optimization. 
Plan-based strategies, by contrast, create a full schedule before execution begins, leveraging a broader decision horizon to achieve better results for objectives like makespan.
In this section, we focus on plan-based strategies as they benefit more from accurate runtime estimates.

A common objective in task scheduling is minimizing the makespan, which serves as a proxy for the total execution time of the workflow~\cite{liu2018survey}. 
However, other objectives and constraints, such as fault tolerance~\cite{benoit2021resilient}, throughput or latency~\cite{benoit2007multi}, and energy efficiency~(see Section~\ref{sec:energycarbonefficieny}), have also been studied. 
These objectives often conflict with each other; for instance, reducing energy consumption may lead to longer execution times. 
Addressing such trade-offs often requires multi-objective optimization, adding complexity to the scheduling problem.

The computational complexity of finding optimal schedules for large scientific workflows, particularly on heterogeneous or distributed platforms, has motivated extensive research into heuristic approaches. 
These methods provide practical solutions by approximating optimal schedules. 
This section focuses on heuristic approaches, summarizing selected plan-based works, highlighting their contributions to the field of workflow task scheduling.

\paragraph{Selected Works:}

Creating a plan-based schedule is NP-hard, even on homogeneous platforms~\cite{garey1979computers}, motivating the use of heuristic optimization methods in practice. 
Among these, two common approaches are list scheduling and partitioning-based heuristics. 
List scheduling algorithms can vary in their approach, with many employing greedy strategies, while partitioning-based methods divide the workflow into blocks to reduce the search space.

HEFT~\cite{topcuoglu2002performance} is a seminal list-scheduling algorithm, which has been extended in various ways to address different scheduling problem formulations, including PEFT~\cite{arabnejad2013list}. 
Further extensions include the work of Shi and Dongarra~\cite{SHI2006665}, as well as dynamic variants like DVR HEFT~\cite{SANDOKJI2019482} and P-HEFT~\cite{pheft}. 
Additionally, cloud-oriented adaptations, such as E-HEFT~\cite{samadi2018eheft}, demonstrate the flexibility of this approach across diverse execution environments.

Partitioning-based heuristic schedulers, including dagP~\cite{Ozkaya19-IPDPS}, DagHetPart~\cite{kulagina2024mapping}, and the framework by Viil and Srirama~\cite{viil2018framework}, are particularly beneficial for large-scale workflows. 
By grouping tasks into blocks and assigning these blocks to nodes, these methods reduce the computational complexity compared to scheduling individual tasks.

Dealing with multiple objectives adds further complexity. 
One common approach is to combine multiple objectives into a single one, as demonstrated by Liu et al.~\cite{liu2016multi}. 
Alternatively, Pareto-optimal solutions can be sought, as done by Su et al.~\cite{su2013cost}. 
For example, Bathie et al.~\cite{bathie2021dynamic} propose a dynamic ILP-based scheduling algorithm that accounts for memory constraints. 
However, these exact approaches are computationally expensive and thus mostly limited to smaller workflows.

Another crucial challenge is the impact of imprecise task runtime estimations. 
Chirkin et al.~\cite{CHIRKIN2017376} explore stochastic scheduling methods to mitigate such inaccuracies, emphasizing the importance of preparation steps in improving scheduling performance. 
Additionally, practical systems have begun integrating monitoring tools that track resource consumption and refine scheduling decisions over time~\cite{bader2021tarema,liu2017scientific,lehmann2023workflow}. 
These systems highlight the growing importance of dynamic, data-driven scheduling in real-world applications.

\paragraph{Conclusion:}

\begin{itemize}
    \item Scheduling algorithms differ in their objective functions and solution approaches. Among these, minimizing the overall makespan of the workflow remains the most commonly pursued objective in workflow scheduling.
    \item Most scheduling approaches depend on accurate estimates of task resource requirements. Inaccurate estimates can lead to suboptimal schedules, impacting both performance and efficiency.
    \item Plan-based techniques, which often achieve better makespans due to their broader decision horizon, are particularly sensitive to errors in task resource predictions.
    \item To address uncertainties in task resource estimations, practical systems increasingly integrate adaptive techniques, such as runtime monitoring and feedback mechanisms, to refine scheduling decisions dynamically.
\end{itemize}

\subsection{Energy Efficiency and Carbon Awareness}
\label{sec:energycarbonefficieny}

\paragraph{Introduction:}

Executing workflows more energy-efficiently reduces their energy consumption.
This can often be achieved by right-sizing, only allocating for example the amount of memory required by workflow tasks.
Similarly, scheduling can reduce energy consumption by ensuring that cluster resources are utilized well by task instances and little energy goes to idling resources.
In contrast, improving the carbon efficiency of a workflow reduces its carbon footprint without necessarily reducing its energy consumption.
This is possible by aligning a workflow's execution and, therefore, its load and energy consumption with the availability of low-carbon energy from variable renewable sources such as wind and solar.
All these optimizations have in common that knowledge of the runtime and resource utilization of scientific workflow tasks is useful, so that they directly benefit from accurate performance predictions.

In the past, research has predominantly focused on improving the energy efficiency of workflows \cite{durilloMultiobjectiveEnergyefficientWorkflow2014b, durilloParetoTradeoffScheduling2015, reddyMultiObjectiveBasedScheduling2023a}.
More recently, however, research has started to optimize specifically for carbon awareness of delay-tolerant batch processing workloads~\cite{wiesnerLetWaitAwhile2021e,radovanovicCarbonAwareComputingDatacenters2023a, linAdaptingDatacenterCapacity2023d}.
In this section, we first summarize selected related work that improves the energy efficiency and carbon awareness of workflows, before we draw a conclusion.

\paragraph{Selected Works:} 

A range of methods has been proposed to improve energy efficiency and carbon awareness for workflows. 
Below, we summarize key contributions in both areas.

GreenHEFT~\cite{durilloMultiobjectiveEnergyefficientWorkflow2014b} and MOHEFT~\cite{durilloParetoTradeoffScheduling2015} are widely recognized algorithms for energy-efficient workflow scheduling. 
GreenHEFT assigns tasks to nodes based on the lowest estimated energy consumption, while MOHEFT considers a trade-off between predicted runtime and energy consumption.
Both methods rely on performance predictions using models trained on historical data.

Reddy et al.~\cite{reddyMultiObjectiveBasedScheduling2023a} introduce a scheduling framework that clusters tasks based on their runtime and assigns clusters to virtual machines (VMs) according to predicted resource availability. 
Their simulation results show reductions in resource consumption, workflow makespan, and energy usage. 
However, as task clustering depends on runtime estimates, the approach is sensitive to inaccuracies in these predictions.

Wen et al.~\cite{wenRunningIndustrialWorkflow2021b} propose an adaptive scheduling approach for industrial workflows using a genetic algorithm. 
The method minimizes monetary costs and maximizes the use of low-carbon energy by scheduling tasks across geographically distributed data centers. 
This approach demonstrates the potential for reducing carbon emissions while maintaining low additional execution costs.

More recent carbon-aware execution methods align workloads with the variable availability of low-carbon energy in a single location, leveraging temporal load shifting and resource scaling. This goes beyond the previous approach, which migrates workloads to locations where more renewable energy is available, as such migration is not always possible due to security/privacy constraints and data transmission costs. 

Wiesner et al.~\cite{wiesnerLetWaitAwhile2021e} examine the potential of shifting the execution of delay-tolerant workloads to times when more low-carbon energy is available to reduce carbon emissions. The simulations assume that the duration of tasks and iterations is known prior to executing jobs. 
In our work~\cite{bader2024lotaru}, we address this assumption by using a task runtime prediction model for carbon-aware scheduling and, thereby, demonstrate the relevance of accurate runtime predictions in reducing carbon emissions.

Souza et al.~\cite{souza2023ecovisor} move beyond simulations by experimentally evaluating carbon savings from simple carbon-aware execution mechanisms, such as applying them to workflows like BLAST. 
These experiments assume that task runtimes are known in advance, which may not reflect real-world conditions.

Hanafy et~al.~\cite{hanafyCarbonScalerLeveragingCloud2023c} develop an algorithm to minimize a workload’s carbon emissions through carbon-aware scaling. This varies the resource usage of a given batch processing application according to variations in the carbon intensity of grid energy, allocating more resources when more low-carbon energy is available. The approach assumes knowledge of the application's runtime and relies on a measured scaling factor obtained through short profiling runs. As such, the method would be affected by inaccurate runtime estimates and also assumes that the application’s scaling behavior is constant even for long-running applications.

We note that none of these recent methods are specific to scientific workflows, despite workflows being extraordinarily suited to carbon-aware scheduling and scaling, as they are commonly significantly delay tolerant, efficiently interruptible, and highly scalable in terms of allocations for both individual tasks and entire workflows, leading to a substantial potential for carbon-optimized execution~\cite{westCCGrid2025}.

\paragraph{Conclusion:}
\begin{itemize}
    \item State-of-the-art methods are able to reduce resource consumption, workflow makespan, energy consumption, and carbon emissions using knowledge of workflow task resource requirements.
    \item However, these methods assume that accurate information on the expected task runtime and resource usage are available, which may not be the case in real-world systems. 
    \item Furthermore, there are few methods that are workflow-specific and also few that simultaneously aim at energy and carbon efficiency by, for example, first targeting high resource utilization and then aligning execution with the availability of low-carbon energy. 
\end{itemize}

\subsection{Cost Optimization and Prediction}

\paragraph{Introduction:}
Cost prediction and optimization for large-scale workflows is crucial for scientists and enterprises, as they regularly work with limited budgets~\cite{rosa2018cost,rosa2021computational}.
Workflow task performance predictions are a prerequisite for this, providing key information such as expected task runtimes and resource requirements.
Costs can be monetary if the target infrastructure is a cloud environment, but they can also be execution slots on shared clusters that are not directly monetary.
As cost optimization and prediction is a broad topic, we provide representative examples covering workflow cost prediction for cloud services and minimizing cost under deadline constraints. 
In this section, we summarize selected related work on predicting and optimizing the cost of executing a workflow.

\paragraph{Selected Works:}
Rosa~et~al.~\cite{rosa2018cost,rosa2021computational} propose provisioning services for cloud federations that can report the cost of workflow execution beforehand.
This is achieved by predicting execution times using a multiple linear regression model based on historical data.
Validated with two real-world bioinformatic workflows, this approach not only ensures accuracy but also empowers users to optimize the cost and efficiency of utilizing federated cloud resources for scientific workflows.

Alkhanak et al.~\cite{alkhanak2016cost} provide an overview of scheduling methods for scientific workflows, emphasizing cost optimization and offering a taxonomy of cost-centric metrics.
The authors note that the cost of executing a workflow on an infrastructure and the time it takes to do so are inversely proportional, as faster nodes tend to be more expensive, but these can be conflicting optimizations.
As a result, the scheduling mechanism must balance the cost of execution and the timing of the workflow, both of which are influenced by inaccuracies in resource predictions.

Malawski~et~al.~\cite{malawski2015scheduling} propose a method to minimize the cost of executing scientific workflows on cloud infrastructures. 
Their model employs mathematical programming languages to formulate a mixed integer programming problem, utilizing knowledge about task runtimes as one of the inputs. The models are evaluated with synthetic workflows and the real-world Montage workflow running on AWS, demonstrating optimization for cost-efficiency.

Zhou~et~al.~\cite{zhou2019cost} propose two approaches that focus on minimizing the cost of scheduling workflows while adhering to deadline constraints, as well as an approach that aims to optimize cost and makespan at the same time.
Their approaches use genetic algorithms, including chromosome encoding, evaluation and selection, and crossover and mutation.
Their simulated evaluation shows that their approach significantly reduces monetary costs compared to existing algorithms under the same deadline constraints.

\paragraph{Conclusion:}
\begin{itemize}
    \item State-of-the-art methods from the literature can achieve significant cost optimizations in cloud environments and accurately predict execution costs using detailed information about the resource usage of workflow tasks.
    \item However, these methods typically assume accurate resource prediction, an assumption that may not hold in real-world systems and leaves room for research that accounts for these uncertainties.
    \item State-of-the-art methods predominantly target cost prediction and optimization in cloud environments, with limited focus on shared cluster environments, where costs are measured in execution slots or time rather than monetary terms.
    
\end{itemize}

\section{Summary}
\label{sec:conclusion}

In this chapter, we have presented an overview of state-of-the-art runtime and memory prediction methods for workflow tasks.
First, we identified key characteristics of task performance prediction methods.
Second, we described and compared state-of-the-art methods in detail.
Third, we explained how such runtime and memory prediction methods are useful for advanced resource management methods such as workflow scheduling, energy-efficient and carbon-aware workflow execution, and workflow cost prediction and optimization.

Our overview of state-of-the-art runtime prediction methods has shown that many methods account for hardware heterogeneity, but none targets task execution on GPUs, a class of devices that is becoming widely used for AI/ML applications.
Meanwhile, heterogeneity is not considered for any of the memory prediction methods.
We assume that this is due to less variability in memory consumption assuming the same operating system and architecture, but it leaves room for further research if such assumptions are not met.
Furthermore, most methods for runtime and memory prediction have only been evaluated via simulations based on workflow execution traces.
Finally, we have observed that the majority of publications do not publish their code, impeding reproducibility and reuse. 

\printbibliography{myChapter}

%% file: main.bbl
\newcommand{\etalchar}[1]{$^{#1}$}
\begin{thebibliography}{RVTBJ{\etalchar{+}}17}

\bibitem[AB13]{arabnejad2013list}
Hamid Arabnejad and Jorge~G Barbosa.
\newblock List scheduling algorithm for heterogeneous systems by an optimistic
  cost table.
\newblock {\em IEEE transactions on parallel and distributed systems},
  25(3):682--694, 2013.

\bibitem[ABC{\etalchar{+}}20]{ahn2020flux}
Dong~H Ahn, Ned Bass, Albert Chu, Jim Garlick, Mark Grondona, Stephen Herbein,
  Helgi~I Ing{\'o}lfsson, Joseph Koning, Tapasya Patki, Thomas~RW Scogland,
  et~al.
\newblock Flux: Overcoming scheduling challenges for exascale workflows.
\newblock {\em Future Generation Computer Systems}, 110:202--213, 2020.

\bibitem[ALRP16]{alkhanak2016cost}
Ehab~Nabiel Alkhanak, Sai~Peck Lee, Reza Rezaei, and Reza~Meimandi Parizi.
\newblock Cost optimization approaches for scientific workflow scheduling in
  cloud and grid computing: A review, classifications, and open issues.
\newblock {\em Journal of Systems and Software}, 113:1--26, 2016.

\bibitem[BDTK23]{baderPredictDynamicMemoryRequ2023}
Jonathan Bader, Nils Diedrich, Lauritz Thamsen, and Odej Kao.
\newblock Predicting dynamic memory requirements for scientific workflow tasks.
\newblock In {\em 2023 IEEE International Conference on Big Data (Big Data)},
  2023.

\bibitem[BLFP{\etalchar{+}}21]{benoit2021resilient}
Anne Benoit, Valentin Le~F{\`e}vre, Lucas Perotin, Padma Raghavan, Yves Robert,
  and Hongyang Sun.
\newblock Resilient scheduling of moldable parallel jobs to cope with silent
  errors.
\newblock {\em IEEE Transactions on Computers}, 71(7):1696--1710, 2021.

\bibitem[BLSHS05]{bailey2005user}
Cynthia Bailey~Lee, Yael Schwartzman, Jennifer Hardy, and Allan Snavely.
\newblock Are user runtime estimates inherently inaccurate?
\newblock In {\em Job Scheduling Strategies for Parallel Processing: 10th
  International Workshop, JSSPP 2004, New York, NY, USA, June 13, 2004. Revised
  Selected Papers 10}, pages 253--263. Springer, 2005.

\bibitem[BLT{\etalchar{+}}22]{bader2022lotaru}
Jonathan Bader, Fabian Lehmann, Lauritz Thamsen, Jonathan Will, Ulf Leser, and
  Odej Kao.
\newblock Lotaru: Locally estimating runtimes of scientific workflow tasks in
  heterogeneous clusters.
\newblock In {\em Proceedings of the 34th International Conference on
  Scientific and Statistical Database Management}, pages 1--12, 2022.

\bibitem[BLT{\etalchar{+}}24a]{bader2024lotaru}
Jonathan Bader, Fabian Lehmann, Lauritz Thamsen, Ulf Leser, and Odej Kao.
\newblock Lotaru: Locally predicting workflow task runtimes for resource
  management on heterogeneous infrastructures.
\newblock {\em Future Generation Computer Systems}, 150:171--185, 2024.

\bibitem[BLT{\etalchar{+}}24b]{bader2024ks+}
Jonathan Bader, Ansgar L{\"o}{\ss}er, Lauritz Thamsen, Bj{\"o}rn Scheuermann,
  and Odej Kao.
\newblock Ks+: Predicting workflow task memory usage over time.
\newblock In {\em 2024 IEEE 20th International Conference on e-Science
  (e-Science)}, pages 1--7. IEEE, 2024.

\bibitem[BM11]{pheft}
Jorge~G Barbosa and Belmiro Moreira.
\newblock Dynamic scheduling of a batch of parallel task jobs on heterogeneous
  clusters.
\newblock {\em Parallel computing}, 37(8), 2011.

\bibitem[BMRT21]{bathie2021dynamic}
Gabriel Bathie, Loris Marchal, Yves Robert, and Samuel Thibault.
\newblock Dynamic dag scheduling under memory constraints for shared-memory
  platforms.
\newblock {\em International Journal of Networking and Computing},
  11(1):27--49, 2021.

\bibitem[BRSR07]{benoit2007multi}
Anne Benoit, Veronika Rehn-Sonigo, and Yves Robert.
\newblock Multi-criteria scheduling of pipeline workflows.
\newblock In {\em 2007 IEEE International Conference on Cluster Computing},
  pages 515--524. IEEE, 2007.

\bibitem[BSL{\etalchar{+}}24]{bader2024Sizey}
Jonathan Bader, Fabian Skalski, Fabian Lehmann, Dominik Scheinert, Jonathan
  Will, Lauritz Thamsen, and Odej Kao.
\newblock Sizey: Memory-efficient execution of scientific workflow tasks.
\newblock In {\em 2024 IEEE International Conference on Cluster Computing
  (CLUSTER)}, 2024.

\bibitem[BTK{\etalchar{+}}21]{bader2021tarema}
Jonathan Bader, Lauritz Thamsen, Svetlana Kulagina, Jonathan Will, Henning
  Meyerhenke, and Odej Kao.
\newblock {Tarema: Adaptive Resource Allocation for Scalable Scientific
  Workflows in Heterogeneous Clusters}.
\newblock In {\em BigData}. IEEE, 2021.

\bibitem[BZBK22]{bader2022leveraging}
Jonathan Bader, Nicolas Zunker, Soeren Becker, and Odej Kao.
\newblock Leveraging reinforcement learning for task resource allocation in
  scientific workflows.
\newblock In {\em 2022 IEEE International Conference on Big Data (Big Data)}.
  IEEE, 2022.

\bibitem[CBK{\etalchar{+}}17]{CHIRKIN2017376}
Artem~M. Chirkin, Adam~S.Z. Belloum, Sergey~V. Kovalchuk, Marc~X. Makkes,
  Mikhail~A. Melnik, Alexander~A. Visheratin, and Denis~A. Nasonov.
\newblock Execution time estimation for workflow scheduling.
\newblock {\em Future Generation Computer Systems}, 75:376--387, 2017.

\bibitem[DNP14]{durilloMultiobjectiveEnergyefficientWorkflow2014b}
Juan~J. Durillo, Vlad Nae, and Radu Prodan.
\newblock Multi-objective energy-efficient workflow scheduling using list-based
  heuristics.
\newblock {\em Future Generation Computer Systems}, 36:221--236, July 2014.

\bibitem[DPB15]{durilloParetoTradeoffScheduling2015}
Juan~J. Durillo, Radu Prodan, and Jorge~G. Barbosa.
\newblock Pareto tradeoff scheduling of workflows on federated commercial
  {{Clouds}}.
\newblock {\em Simulation Modelling Practice and Theory}, 58:95--111, November
  2015.

\bibitem[DSJD{\etalchar{+}}13]{da2013toward}
Rafael~Ferreira Da~Silva, Gideon Juve, Ewa Deelman, Tristan Glatard,
  Fr{\'e}d{\'e}ric Desprez, Douglas Thain, Benjam{\'\i}n Tovar, and Miron
  Livny.
\newblock Toward fine-grained online task characteristics estimation in
  scientific workflows.
\newblock In {\em Proceedings of the 8th Workshop on Workflows in Support of
  Large-Scale Science}, pages 58--67, 2013.

\bibitem[DSJR{\etalchar{+}}15]{da2015online}
Rafael~Ferreira Da~Silva, Gideon Juve, Mats Rynge, Ewa Deelman, and Miron
  Livny.
\newblock Online task resource consumption prediction for scientific workflows.
\newblock {\em Parallel Processing Letters}, 25(03), 2015.

\bibitem[EPF{\etalchar{+}}20]{ewels2020nf}
Philip~A Ewels, Alexander Peltzer, Sven Fillinger, Harshil Patel, Johannes
  Alneberg, Andreas Wilm, Maxime~Ulysse Garcia, Paolo Di~Tommaso, and Sven
  Nahnsen.
\newblock The nf-core framework for community-curated bioinformatics pipelines.
\newblock {\em Nature biotechnology}, 38(3):276--278, 2020.

\bibitem[GJ79]{garey1979computers}
M.~R. Garey and D.~S. Johnson.
\newblock {\em Computers and Intractability: A Guide to the Theory of
  NP-Completeness (Series of Books in the Mathematical Sciences)}.
\newblock W. H. Freeman, 1979.

\bibitem[HCP{\etalchar{+}}23]{huang2023cloudprophet}
Darong Huang, Luis Costero, Ali Pahlevan, Marina Zapater, and David Atienza.
\newblock Cloudprophet: A machine learning-based performance prediction for
  public clouds.
\newblock {\em arXiv preprint arXiv:2309.16333}, 2023.

\bibitem[HCTY{\etalchar{+}}12]{hirales2012multiple}
Ad{\'a}n Hirales-Carbajal, Andrei Tchernykh, Ramin Yahyapour, Jos{\'e}~Luis
  Gonz{\'a}lez-Garc{\'\i}a, Thomas R{\"o}blitz, and Juan~Manuel
  Ram{\'\i}rez-Alcaraz.
\newblock Multiple workflow scheduling strategies with user run time estimates
  on a grid.
\newblock {\em Journal of Grid Computing}, 10:325--346, 2012.

\bibitem[HLB{\etalchar{+}}23]{hanafyCarbonScalerLeveragingCloud2023c}
Walid~A. Hanafy, Qianlin Liang, Noman Bashir, David Irwin, and Prashant Shenoy.
\newblock {{CarbonScaler}}: {{Leveraging Cloud Workload Elasticity}} for
  {{Optimizing Carbon-Efficiency}}.
\newblock {\em Proc. ACM Meas. Anal. Comput. Syst.}, 7(3):57:1--57:28, December
  2023.

\bibitem[HRB18]{hilman2018task}
Muhammad~Hafizhuddin Hilman, Maria~Alejandra Rodriguez, and Rajkumar Buyya.
\newblock Task runtime prediction in scientific workflows using an online
  incremental learning approach.
\newblock In {\em 2018 IEEE/ACM 11th International Conference on Utility and
  Cloud Computing (UCC)}, pages 93--102. IEEE, 2018.

\bibitem[IE18]{ilyushkin2018impact}
Alexey Ilyushkin and Dick Epema.
\newblock The impact of task runtime estimate accuracy on scheduling workloads
  of workflows.
\newblock In {\em 2018 18th IEEE/ACM International Symposium on Cluster, Cloud
  and Grid Computing (CCGRID)}, pages 331--341. IEEE, 2018.

\bibitem[KMB24]{kulagina2024mapping}
Svetlana Kulagina, Henning Meyerhenke, and Anne Benoit.
\newblock Mapping large memory-constrained workflows onto heterogeneous
  platforms.
\newblock In {\em 2024 53rd International Conference on Parallel Processing
  (ICPP)}, 2024.

\bibitem[LBDM{\etalchar{+}}24]{lehmann2024ponder}
Fabian Lehmann, Jonathan Bader, Ninon De~Mecquenem, Xing Wang, Vasilis
  Bountris, Florian Friederici, Ulf Leser, and Lauritz Thamsen.
\newblock {Ponder: Online Prediction of Task Memory Requirements for Scientific
  Workflows}.
\newblock In {\em {Proceedings of the 2024 IEEE 20th International Conference
  on e-Science (e-Science)}}, e-Science '24. IEEE, September 2024.

\bibitem[LBT{\etalchar{+}}23]{lehmann2023workflow}
Fabian Lehmann, Jonathan Bader, Friedrich Tschirpke, Lauritz Thamsen, and Ulf
  Leser.
\newblock How workflow engines should talk to resource managers: A proposal for
  a common workflow scheduling interface.
\newblock In {\em 2023 IEEE/ACM 23rd International Symposium on Cluster, Cloud
  and Internet Computing (CCGrid)}, pages 166--179. IEEE, 2023.

\bibitem[LC23]{linAdaptingDatacenterCapacity2023d}
Liuzixuan Lin and Andrew~A. Chien.
\newblock Adapting {{Datacenter Capacity}} for {{Greener Datacenters}} and
  {{Grid}}.
\newblock In {\em Proceedings of the 14th {{ACM International Conference}} on
  {{Future Energy Systems}}}, pages 200--213, June 2023.

\bibitem[LPV{\etalchar{+}}16]{liu2016multi}
Ji~Liu, Esther Pacitti, Patrick Valduriez, Daniel De~Oliveira, and Marta
  Mattoso.
\newblock Multi-objective scheduling of scientific workflows in multisite
  clouds.
\newblock {\em Future Generation Computer Systems}, 63:76--95, 2016.

\bibitem[LPV18]{liu2018survey}
Ji~Liu, Esther Pacitti, and Patrick Valduriez.
\newblock A survey of scheduling frameworks in big data systems.
\newblock {\em International Journal of Cloud Computing}, 7(2):103--128, 2018.

\bibitem[LPVM17]{liu2017scientific}
Ji~Liu, Esther Pacitti, Patrick Valduriez, and Marta Mattoso.
\newblock Scientific workflow scheduling with provenance data in a multisite
  cloud.
\newblock {\em Transactions on large-scale data-and knowledge-centered systems
  XXXIII}, pages 80--112, 2017.

\bibitem[MF10]{matsunaga2010use}
Andr{\'e}a Matsunaga and Jos{\'e}~AB Fortes.
\newblock On the use of machine learning to predict the time and resources
  consumed by applications.
\newblock In {\em 2010 10th IEEE/ACM International Conference on Cluster, Cloud
  and Grid Computing}, pages 495--504. IEEE, 2010.

\bibitem[MFB{\etalchar{+}}15]{malawski2015scheduling}
Maciej Malawski, Kamil Figiela, Marian Bubak, Ewa Deelman, and Jarek Nabrzyski.
\newblock Scheduling multilevel deadline-constrained scientific workflows on
  clouds based on cost optimization.
\newblock {\em Scientific Programming}, 2015(1):680271, 2015.

\bibitem[MFP13]{malik2013execution}
Muhammad~Junaid Malik, Thomas Fahringer, and Radu Prodan.
\newblock Execution time prediction for grid infrastructures based on runtime
  provenance data.
\newblock In {\em Proceedings of the 8th Workshop on Workflows in Support of
  Large-Scale Science}, pages 48--57, 2013.

\bibitem[MPZ{\etalchar{+}}24]{mohammadi2024optimizing}
Somayeh Mohammadi, Latif PourKarimi, Manuel Zsch{\"a}bitz, Tristan Aretz, Ninon
  De~Mecquenem, Ulf Leser, and Knut Reinert.
\newblock Optimizing job/task granularity for metagenomic workflows in
  heterogeneous cluster infrastructures.
\newblock In {\em EDBT/ICDT Workshops}, 2024.

\bibitem[NAM{\etalchar{+}}17]{nadeem2017modeling}
Farrukh Nadeem, Daniyal Alghazzawi, Abdulfattah Mashat, Khalid Fakeeh, Abdullah
  Almalaise, and Hani Hagras.
\newblock Modeling and predicting execution time of scientific workflows in the
  grid using radial basis function neural network.
\newblock {\em Cluster Computing}, 20(3):2805--2819, 2017.

\bibitem[OBU{\etalchar{+}}19]{Ozkaya19-IPDPS}
M.~Yusuf \"Ozkaya, Anne Benoit, Bora U{\c{c}}ar, Julien Herrmann, and
  {\"{U}}mit~V. {\c{C}}ataly{\"{u}}rek.
\newblock A scalable clustering-based task scheduler for homogeneous processors
  using dag partitioning.
\newblock In {\em 33rd IEEE International Parallel and Distributed Processing
  Symposium}, May 2019.

\bibitem[PDF17]{pham2017predicting}
Thanh-Phuong Pham, Juan~J Durillo, and Thomas Fahringer.
\newblock Predicting workflow task execution time in the cloud using a
  two-stage machine learning approach.
\newblock {\em IEEE Transactions on Cloud Computing}, 8(1):256--268, 2017.

\bibitem[PWCT21]{phung2021not}
Thanh~Son Phung, Logan Ward, Kyle Chard, and Douglas Thain.
\newblock Not all tasks are created equal: Adaptive resource allocation for
  heterogeneous tasks in dynamic workflows.
\newblock In {\em 2021 IEEE Workshop on Workflows in Support of Large-Scale
  Science (WORKS)}, pages 17--24. IEEE, 2021.

\bibitem[RAM18]{rosa2018cost}
Michel~JF Rosa, Alet{\'e}ia~PF Ara{\"u}jo, and Felipe~LS Mendes.
\newblock Cost and time prediction for efficient execution of bioinformatics
  workflows in federated cloud.
\newblock In {\em 2018 IEEE International Conference on Bioinformatics and
  Biomedicine (BIBM)}, pages 1703--1710. IEEE, 2018.

\bibitem[RKS{\etalchar{+}}23]{radovanovicCarbonAwareComputingDatacenters2023a}
Ana Radovanovi{\'c}, Ross Koningstein, Ian Schneider, Bokan Chen, Alexandre
  Duarte, Binz Roy, Diyue Xiao, Maya Haridasan, Patrick Hung, Nick Care, Saurav
  Talukdar, Eric Mullen, Kendal Smith, MariEllen Cottman, and Walfredo Cirne.
\newblock Carbon-{{Aware Computing}} for {{Datacenters}}.
\newblock {\em IEEE Transactions on Power Systems}, 38(2):1270--1280, March
  2023.

\bibitem[RR23]{reddyMultiObjectiveBasedScheduling2023a}
Pillareddy~Vamsheedhar Reddy and Karri~Ganesh Reddy.
\newblock A {{Multi-Objective Based Scheduling Framework}} for {{Effective
  Resource Utilization}} in {{Cloud Computing}}.
\newblock {\em IEEE Access}, 11:37178--37193, 2023.

\bibitem[RRHA21]{rosa2021computational}
Michel~JF Rosa, C{\'e}lia~Ghedini Ralha, Maristela Holanda, and Aleteia~PF
  Araujo.
\newblock Computational resource and cost prediction service for scientific
  workflows in federated clouds.
\newblock {\em Future Generation Computer Systems}, 125:844--858, 2021.

\bibitem[RVTBJ{\etalchar{+}}17]{ramirez2017adaptive}
Raul Ram{\'\i}rez-Velarde, Andrei Tchernykh, Carlos Barba-Jimenez, Ad{\'a}n
  Hirales-Carbajal, and Juan Nolazco-Flores.
\newblock Adaptive resource allocation with job runtime uncertainty.
\newblock {\em Journal of Grid Computing}, 15:415--434, 2017.

\bibitem[SBM{\etalchar{+}}23]{souza2023ecovisor}
Abel Souza, Noman Bashir, Jorge Murillo, Walid Hanafy, Qianlin Liang, David
  Irwin, and Prashant Shenoy.
\newblock Ecovisor: A virtual energy system for carbon-efficient applications.
\newblock In {\em Proceedings of the 28th ACM International Conference on
  Architectural Support for Programming Languages and Operating Systems, Volume
  2}, ASPLOS 2023, page 252–265. ACM, 2023.

\bibitem[SD06]{SHI2006665}
Zhiao Shi and Jack~J. Dongarra.
\newblock Scheduling workflow applications on processors with different
  capabilities.
\newblock {\em Future Generation Computer Systems}, 22(6):665--675, 2006.

\bibitem[SE19]{SANDOKJI2019482}
Suhelah Sandokji and Fathy Eassa.
\newblock Dynamic variant rank heft task scheduling algorithm toward exascle
  computing.
\newblock {\em Procedia Computer Science}, 163:482--493, 2019.
\newblock 16th Learning and Technology Conference 2019Artificial Intelligence
  and Machine Learning: Embedding the Intelligence.

\bibitem[SLH{\etalchar{+}}13]{su2013cost}
Sen Su, Jian Li, Qingjia Huang, Xiao Huang, Kai Shuang, and Jie Wang.
\newblock Cost-efficient task scheduling for executing large programs in the
  cloud.
\newblock {\em Parallel Computing}, 39(4-5):177--188, 2013.

\bibitem[SZT18]{samadi2018eheft}
Yassir Samadi, Mostapha Zbakh, and Claude Tadonki.
\newblock E-heft: Enhancement heterogeneous earliest finish time algorithm for
  task scheduling based on load balancing in cloud computing.
\newblock In {\em 2018 International Conference on High Performance Computing
  \& Simulation (HPCS)}, pages 601--609, 2018.

\bibitem[TdSJ{\etalchar{+}}17]{tovar2017job}
Benjamin Tovar, Rafael~Ferreira da~Silva, Gideon Juve, Ewa Deelman, William
  Allcock, Douglas Thain, and Miron Livny.
\newblock A job sizing strategy for high-throughput scientific workflows.
\newblock {\em IEEE Transactions on Parallel and Distributed Systems}, 29(2),
  2017.

\bibitem[THW02a]{topcuoglu2002performance}
Haluk Topcuoglu, Salim Hariri, and Min-You Wu.
\newblock Performance-effective and low-complexity task scheduling for
  heterogeneous computing.
\newblock {\em IEEE transactions on parallel and distributed systems},
  13(3):260--274, 2002.

\bibitem[THW02b]{heft}
Haluk Topcuoglu, Salim Hariri, and Min-you Wu.
\newblock Performance-effective and low-complexity task scheduling for
  heterogeneous computing.
\newblock {\em IEEE transactions on parallel and distributed systems}, 13(3),
  2002.

\bibitem[TLM{\etalchar{+}}22]{tovar2022dynamic}
Ben Tovar, Ben Lyons, Kelci Mohrman, Barry Sly-Delgado, Kevin Lannon, and
  Douglas Thain.
\newblock Dynamic task shaping for high throughput data analysis applications
  in high energy physics.
\newblock In {\em 2022 IEEE International Parallel and Distributed Processing
  Symposium (IPDPS)}. IEEE, 2022.

\bibitem[VS18]{viil2018framework}
Jaagup Viil and Satish~Narayana Srirama.
\newblock Framework for automated partitioning and execution of scientific
  workflows in the cloud.
\newblock {\em The Journal of Supercomputing}, 74:2656--2683, 2018.

\bibitem[WBGL19]{witt2019predictive}
Carl Witt, Marc Bux, Wladislaw Gusew, and Ulf Leser.
\newblock Predictive performance modeling for distributed batch processing
  using black box monitoring and machine learning.
\newblock {\em Information Systems}, 82:33--52, 2019.

\bibitem[WBS{\etalchar{+}}21]{wiesnerLetWaitAwhile2021e}
Philipp Wiesner, Ilja Behnke, Dominik Scheinert, Kordian Gontarska, and Lauritz
  Thamsen.
\newblock Let's wait awhile: How temporal workload shifting can reduce carbon
  emissions in the cloud.
\newblock In {\em Proceedings of the 22nd {{International Middleware
  Conference}}}, Middleware '21, pages 260--272. ACM, 2021.

\bibitem[WGA{\etalchar{+}}21]{wenRunningIndustrialWorkflow2021b}
Zhenyu Wen, Saurabh Garg, Gagangeet~Singh Aujla, Khaled Alwasel, Deepak Puthal,
  Schahram Dustdar, Albert~Y. Zomaya, and Rajiv Ranjan.
\newblock Running {{Industrial Workflow Applications}} in a {{Software-Defined
  Multicloud Environment Using Green Energy Aware Scheduling Algorithm}}.
\newblock {\em IEEE Transactions on Industrial Informatics}, 17(8):5645--5656,
  August 2021.

\bibitem[WLB{\etalchar{+}}25]{westCCGrid2025}
Kathleen West, Fabian Lehmann, Vasilis Bountris, Ulf Leser, Yehia Elkhatib, and
  Lauritz Thamsen.
\newblock {Exploring the Potential of Carbon-Aware Execution for Scientific
  Workflows}.
\newblock In {\em 2025 {{IEEE}} 25th International Symposium on Cluster, Cloud
  and Internet Computing ({{CCGrid}})}. IEEE, 2025.

\bibitem[WVM96]{weske1996scientific}
Mathias Weske, Gottfried Vossen, and Claudia~Bauzer Medeiros.
\newblock {\em Scientific workflow management: WASA architecture and
  applications}.
\newblock Citeseer, 1996.

\bibitem[WvSL19]{witt2019learning}
Carl Witt, Jakob van Santen, and Ulf Leser.
\newblock Learning low-wastage memory allocations for scientific workflows at
  icecube.
\newblock In {\em 2019 International Conference on High Performance Computing
  \& Simulation (HPCS)}. IEEE, 2019.

\bibitem[WWL19]{witt2019feedback}
Carl Witt, Dennis Wagner, and Ulf Leser.
\newblock Feedback-based resource allocation for batch scheduling of scientific
  workflows.
\newblock In {\em 2019 International Conference on High Performance Computing
  \& Simulation (HPCS)}. IEEE, 2019.

\bibitem[ZWC{\etalchar{+}}19]{zhou2019cost}
Junlong Zhou, Tian Wang, Peijin Cong, Pingping Lu, Tongquan Wei, and Mingsong
  Chen.
\newblock Cost and makespan-aware workflow scheduling in hybrid clouds.
\newblock {\em Journal of Systems Architecture}, 100:101631, 2019.

\end{thebibliography}
